\def\pg{\mbox{PG~0122+200}}
\def\pp{\mbox{PG~1159$-$035}}
\def\pr{\mbox{PG~2131+066}}
\def\pt{\mbox{PG~1707+427}}
\def\rxj{\mbox{RX~J2117.1+3412}}
\def\v4334{\mbox{V4334 Sgr}}
\def\ngc{\mbox{NGC 1501}}

\documentclass{aa}

\usepackage{graphicx}
\usepackage{txfonts}
        
\begin{document}

\title{Asteroseismology of hot pre-white dwarf stars: the case of 
the DOV stars \pr\ and \pt, and the PNNV star \ngc}

\author{A. H. C\'orsico\inst{1,2,3},
        L. G. Althaus\inst{1,2,3}, 
        M. M. Miller Bertolami\inst{1,2,4}, \and
        E. Garc\'\i a--Berro\inst{5,6}}

\offprints{A. H. C\'orsico}

\institute{Facultad de Ciencias Astron\'omicas y Geof\'{\i}sicas,  
           Universidad  Nacional de La Plata,
           Paseo del  Bosque S/N,  
           (1900) La Plata, 
           Argentina\
           \and
           Instituto de Astrof\'{\i}sica La Plata, 
           IALP, CONICET-UNLP,
           Argentina\
           \and 
           Member of the Carrera del Investigador Cient\'{\i}fico y 
           Tecnol\'ogico, CONICET, 
           Argentina\
           \and
           Fellow of CONICET, Argentina\
           \and
           Departament de F\'\i sica Aplicada, 
           Escola Polit\`ecnica Superior de Castelldefels,
           Universitat Polit\`ecnica de Catalunya,  
           Av. del Canal Ol\'\i mpic, s/n,  
           08860 Castelldefels,  
           Spain\
           \and
           Institute for Space Studies of Catalonia,
           c/Gran Capit\`a 2--4, Edif. Nexus 104,   
           08034  Barcelona,  Spain\\  
\email{acorsico,althaus,mmiller@fcaglp.unlp.edu.ar; garcia@fa.upc.edu} }

\abstract{}
         {We   present  an  asteroseismological   study  on   the  two
          high-gravity pulsating PG1159 (GW Vir or DOV) stars, \pr\ and 
          \pt, and on  the pulsating  [WCE] star  \ngc.  All  of these  
          stars  have been intensively  scrutinized through multi-site 
          observations, so they have  well resolved pulsation spectra.}
         {We compute adiabatic  $g$-mode pulsation  periods  on PG1159
          evolutionary models with stellar masses ranging from $0.530$
          to $0.741 M_{\odot}$.   These  models take  into account  the
          complete   evolution  of   progenitor  stars,   through  the
          thermally  pulsing  AGB phase, and  born-again episode.   We
          constrain  the  stellar  mass  of  \pr, \pt,  and  \ngc\  by
          comparing  the observed period  spacing with  the asymptotic
          period spacing  and with the average of  the computed period
          spacings.  We also employ the individual observed periods 
          in search of representative seismological models  for each
          star.}  
         {We  derive a  stellar mass  of $0.627\,  M_{\odot}$  for \pr,
          $0.597  \, M_{\odot}$  for  \pt, and  $0.571\, M_{\odot}$  for
          \ngc\ from a comparison between the observed period spacings
          and the  computed asymptotic period spacings,  and a stellar
          mass of $0.578\, M_{\odot}$  for \pr, $0.566 \, M_{\odot}$ for
          \pt,  and $0.576  \, M_{\odot}$  for \ngc\  by  comparing the
          observed period  spacings with  the average of  the computed
          period spacings.  We also find, on the basis of a period-fit
          procedure,  asteroseismological  models  representatives  of
          \pr\ and  \pt. These best-fit  models are able  to reproduce
          the observed period patterns  of these stars with an average
          of the period  differences of $\overline{\delta\Pi_i}= 1.57$
          s and $\overline{\delta  \Pi_i}= 1.75$ s, respectively.  The
          best-fit model for \pr\ has an effective temperature $T_{\rm
          eff}= 102\,100$ K, a stellar mass $M_*= 0.589\, M_{\odot}$, a
          surface  gravity $\log  g= 7.63$,  a stellar  luminosity and
          radius     of     $\log(L_*/L_{\odot})     =    1.57$     and
          $\log(R_*/R_{\odot})=  -1.71$,  respectively,  and a  He-rich
          envelope  thickness of  $M_{\rm env}=  1.6  \times 10^{-2}\,
          M_{\odot}$. We derive a seismic  distance $d \sim 830$ pc and
          a parallax $\pi \sim 1.2$  mas.  The best-fit model for \pt,
          on  the other  hand,  has $T_{\rm  eff}=  89\,500$ K,  $M_*=
          0.542\,  M_{\odot}$, $\log  g=  7.53$, $\log(L_*/L_{\odot})  =
          1.40$,  $\log(R_*/R_{\odot})= -1.68$,  and $M_{\rm  env}= 2.5
          \times  10^{-2}\, M_{\odot}$,  and the  seismic  distance and
          parallax  are  $d \sim  730$  pc  and  $\pi \sim  1.4$  mas.
          Finally, we have been unable to find an unambiguous best-fit
          model for \ngc\ on the basis of a period-fit procedure.}
         {This  work closes  our short  series  of asteroseismological
          studies  on pulsating  pre-white dwarf  stars.   Our results
          demonstrate the  usefulness of asteroseismology  for probing
          the internal structure  and evolutionary status of pre-white
          dwarf  stars.  In particular, asteroseismology is able to 
          determine stellar masses of PG1159 stars with an accuracy 
          comparable or even better than spectroscopy.}


\authorrunning{C\'orsico et al.}

\titlerunning{Asteroseismology of \pr, \pt, and \ngc.}

\date{\today}

\maketitle

 
\section{Introduction}
\label{intro}

Pulsating PG1159 stars (also called GW  Vir or DOV stars) are very hot
hydrogen-deficient  post-Asymptotic  Giant  Branch  (AGB)  stars  with
surface layers  rich in helium,  carbon, and oxygen (Werner  \& Herwig
2006) that  exhibit multiperiodic  luminosity variations  with periods
ranging  from 5 to  50 minutes,  attributable to  non-radial pulsation
$g$-modes (see  Winget \&  Kepler 2008 for  a recent  review).  PG1159
stars are thought to be  the evolutionary link between Wolf-Rayet type
central stars of planetary  nebulae and most of the hydrogen-deficient
white dwarfs  (Althaus et  al. 2005).  It  is generally  accepted that
these stars  have their  origin in a  born-again episode induced  by a
post-AGB helium thermal  pulse --- see Iben et  al.  (1983), Herwig et
al.  (1999),  Lawlor \& MacDonald  (2003), and Althaus et  al.  (2005)
for recent references.

Recently, considerable observational effort has been invested to study
pulsating  PG1159  stars. Particularly  noteworthy  are  the works  of
Vauclair et al. (2002) on \rxj, Fu et al.  (2007) on \pg, and Costa et
al.  (2008) and Costa \& Kepler  (2008) on \pp.  These stars have been
monitored through  long-term observations  carried out with  the Whole
Earth  Telescope (Nather  et al.   1990).  On  the  theoretical front,
recent important  progress in the  numerical modeling of  PG1159 stars
(Althaus et  al. 2005;  Miller Bertolami \&  Althaus 2006,  2007a) has
paved the way for unprecedented asteroseismological inferences for the
mentioned  stars (C\'orsico  \& Althaus  2006, C\'orsico  et al.
2007a, 2007b, 2008). The  new generation of PG1159 evolutionary models
of Miller  Bertolami \&  Althaus (2006) is  derived from  the complete
evolutionary history of progenitor stars with different stellar masses
and an  elaborate treatment of  the mixing and  extra-mixing processes
during the core  helium burning and born-again phases.  The success of
these models at explaining  the spread in surface chemical composition
observed in PG1159 stars (Miller Bertolami \& Althaus 2006), the short
born-again times  of \v4334 (Miller  Bertolami \& Althaus  2007b), and
the  location of  the GW  Vir instability  strip in  the  $\log T_{\rm
eff}-\log g$ plane (C\'orsico et al.  2006) renders reliability to the
inferences drawn from individual pulsating PG1159 stars.

Besides the mentioned three well-studied pulsating PG1159 stars, there
exist  two other  variable stars  of this  class that  have  been also
intensively  scrutinized through  the multi-site  observations  of the
WET: \pr\ and  \pt.  In addition, there is a  variable central star of
planetary nebula (PNNV), \ngc, which  has been the subject of a nearly
continuous photometric  coverage from  a global observing  campaign by
Bond  et al.   (1996). We  briefly summarize  the properties  of these
stars below.

\pr\  was discovered  as  a variable  star by  Bond et al. (1984) with 
periods   of  about   414   and   386  s,   along   with  some   other
periodicities. On  the basis of an  augmented set of  periods from WET
data, Kawaler et al.  (1995)  obtained a precise mass determination of
$M_*  = 0.61  \, M_{\odot}$,  a luminosity  of $10\,  L_{\odot}$,  and a
seismological distance  from the Earth of $d=  470$ pc.  Spectroscopic
constraints of Dreizler \& Heber  (1998), on the other hand, gave $M_*
= 0.55\, M_{\odot}$, $T_{\rm eff}  = 95\,000$ K, $39.8\, L_{\odot}$, and
$\log g =  7.5$ for PG 2131+066.  By  using this updated determination
of  the  effective  temperature,  Reed  et  al.   (2000)  refined  the
procedure of Kawaler et al.  (1995) and found $M_*= 0.608\, M_{\odot}$,
$L_*= 26\, L_{\odot}$ and $d= 668$ pc.

\pt\ was discovered to  be a pulsator by  Bond et al. (1984). Dreizler 
\& Heber (1998) obtained $T_{\rm eff}= 85\,000$ K,  $\log g= 7.5$, and 
then $M_*=  0.54\, M_{\odot}$ and  $L_*= 25 \, L_{\odot}$  were inferred
from  their spectroscopic  study.   Recently, Kawaler  et al.   (2004)
reported the analysis of multi-site observations of \pt\ obtained with
WET.  Preliminary  seismic  analysis by  using 7 
independent $\ell=  1$ modes with periods  between 334  and 910 s 
suggest  an  asteroseismological  mass and
luminosity of $0.57 \, M_{\odot}$ and $23 \, L_{\odot}$, respectively.

\ngc\ was  classified  as  a  [WCE]  star,   an  early  low-mass  Wolf 
Rayet-type  PNNV with spectra  dominated by  strong helium  and carbon
emission lines (Werner \& Herwig  2006).  The effective temperature 
and gravity of this star are $T_{\rm eff}= 134\,000$ K and  $\log g= 6.0$
(Werner \& Herwig  2006). The variable nature of \ngc\
was  discovered by  Bond  \&  Ciardullo (1993).   The  star shows  ten
periodicities ranging from 5200 s down to 1154 s, although the largest
amplitude  pulsations occur  between  1154  s and  2000  s.  Based  on
period-spacing data, Bond et al.  (1996) found a stellar mass of $0.53
\pm 0.03\, M_{\odot}$ for
\ngc.

In  this work  we  complete our  small  survey of  asteroseismological
inferences on pulsating PG1159 stars  --- see C\'orsico et al. (2007a,
2007b, 2008) for the previous studies of this series --- by performing
a detailed study of the GW Vir  stars \pr\ and \pt, and the [WCE] star
\ngc.  We employ the same stellar models and numerical tools as in our
previous  works.  In  particular, we  go beyond  the mere  use  of the
observed  period-spacing  data by  performing,  in addition,  detailed
period-to-period fits  on the pulsation  spectrum of these  stars.  In
our approach,  we take full  advantage of the  state-of-the-art PG1159
evolutionary models  developed by Miller Bertolami  \& Althaus (2006).
The paper  is organized as follows.  In  Section \ref{evolutionary} we
briefly    describe    our    PG1159    evolutionary    models.     In
Sect.~\ref{period-spacing} we derive the stellar mass of \pr, \pt, and
\ngc\  by   using   the   observed  period-spacing   data   alone.  In 
Sect.~\ref{fitting} we  infer structural parameters of  these stars by
employing the individual observed  periods.  In this section we derive
asteroseismological   models   representative   of   \pr\   and   \pt\
(\ref{searching}), and  discuss their main  structural and pulsational
characteristics (\ref{char}).   In Sect.~\ref{comparison} we compare 
the results of the present paper with those of the 
asteroseismological study of C\'orsico  \& Althaus (2006) 
(hereinafter CA06). 
Finally, in Sect.~\ref{conclusions} we summarize our main results and 
make some concluding remarks.

\section{Evolutionary models and numerical tools}
\label{evolutionary}

The  pulsation  analysis  presented  in  this work  relies  on  a  new
generation  of stellar  models  that take  into  account the  complete
evolution of PG1159 progenitor stars. Specifically, the stellar models
were extracted  from the evolutionary  calculations recently presented
by  Althaus et  al. (2005),  Miller Bertolami  \& Althaus  (2006), and
C\'orsico et al. (2006), who  computed the complete evolution of model
star  sequences with  initial masses  on the  ZAMS ranging  from  1 to
$3.75\, M_{\odot}$.   All of  the post-AGB evolutionary  sequences were
computed  using  the  {\tt   LPCODE}  evolutionary  code  (Althaus  et
al. 2005) and were followed through the very late thermal pulse (VLTP)
and  the   resulting  born-again  episode  that  gives   rise  to  the
H-deficient, He-, C-, and  O-rich composition characteristic of PG1159
stars.  The masses of the  resulting remnants are 0.530, 0.542, 0.556,
0.565, 0.589, 0.609, 0.664,  and $0.741\, M_{\odot}$. For details about
the  input  physics and  evolutionary  code  used,  and the  numerical
simulations  performed  to obtain  the  PG1159 evolutionary  sequences
employed here, we refer the  interested reader to the works by Althaus
et al. (2005) and Miller Bertolami \& Althaus (2006, 2007a).

We computed  $\ell= 1$ $g$-mode  adiabatic pulsation periods  with the
same numerical code and methods we employed in our previous works, see
C\'orsico \&  Althaus (2006) for  details.  In addition,  we performed
nonadiabatic  computations  with the  help  of  the  code employed  in
C\'orsico et al.  (2006) to evaluate the pulsational  stability of the
asteroseismological models presented in Sect.
\ref{fitting}. We  analyzed about  3000
PG1159 models covering a wide range of effective temperatures ($5.4
\ga  \log(T_{\rm eff}) \ga  4.8$) and  luminosities  ($0 \la \log(L_*/
L_{\odot})  \la 4.2$),  and  a  range of  stellar  masses ($0.530  \leq
M_*/M_{\odot} \leq 0.741$).

\begin{table*}[t]
\begin{center}
\caption{Stellar  masses  for all of the intensively studied pulsating  
         PG 1159 stars, including also one pulsating [WCE] star. All  
         masses are in solar units.}
\begin{tabular}{lcccclc}
\hline 
\hline
\noalign{\smallskip}
Star & $\Delta \Pi_{\ell}^{\rm a}$ &$\overline{\Delta \Pi_{\ell}}$ & Approximate  & Period fit & Pulsations    & Spectroscopy\\
     &                             &                               &  formula     &            & (other works) & \\
\noalign{\smallskip}
\hline
\ngc       & 0.571$^{\rm a}$        & 0.576$^{\rm a}$ & 0.530$^{\rm a}$ & ---             & 0.55$^{\rm j}$ (asymptotic analysis) & 0.56 \\ 
\rxj       & 0.568$^{\rm b}$        & 0.560$^{\rm b}$ & 0.525$^{\rm a}$ & 0.565$^{\rm b}$ & 0.56$^{\rm h}$ (asymptotic analysis) & 0.72 \\ 
\pp        & 0.577--0.585$^{\rm d}$ & 0.561$^{\rm d}$ & 0.570$^{\rm a}$ & 0.565$^{\rm d}$ & 0.59$^{\rm i}$ (asymptotic analysis) & 0.54 \\ 
\pr        & 0.627$^{\rm a}$        & 0.578$^{\rm a}$ & 0.609$^{\rm a}$ & 0.589$^{\rm a}$ & 0.61$^{\rm e}$ (period fit)          & 0.55 \\ 
\pt        & 0.597$^{\rm a}$        & 0.566$^{\rm a}$ & 0.587$^{\rm a}$ & 0.542$^{\rm a}$ & 0.57$^{\rm g}$ (asymptotic analysis) & 0.53 \\ 
\pg        & 0.625$^{\rm c}$        & 0.567$^{\rm c}$ & 0.593$^{\rm a}$ & 0.566$^{\rm c}$ & 0.69$^{\rm f}$ (asymptotic analysis) & 0.53 \\ 
\hline
\hline
\end{tabular} 
\label{tabla-masas}
\end{center}

{\footnotesize  Notes: 
$^{\rm a}$This work.   
$^{\rm b}$C\'orsico et al. (2007a).  
$^{\rm c}$C\'orsico et al. (2007b).  
$^{\rm d}$C\'orsico et al. (2008).  
$^{\rm e}$Reed et al. (2000).  
$^{\rm f}$Fu et al. (2007).
$^{\rm g}$Kawaler et al. (2004).   
$^{\rm h}$Vauclair et al. (2002).
$^{\rm i}$Costa et al. (2008). 
$^{\rm j}$Bond et al. (1996).}
\end{table*}

\section{Mass determination from the observed period spacing}
\label{period-spacing}

\begin{figure}
\centering
\includegraphics[clip,width=250pt]{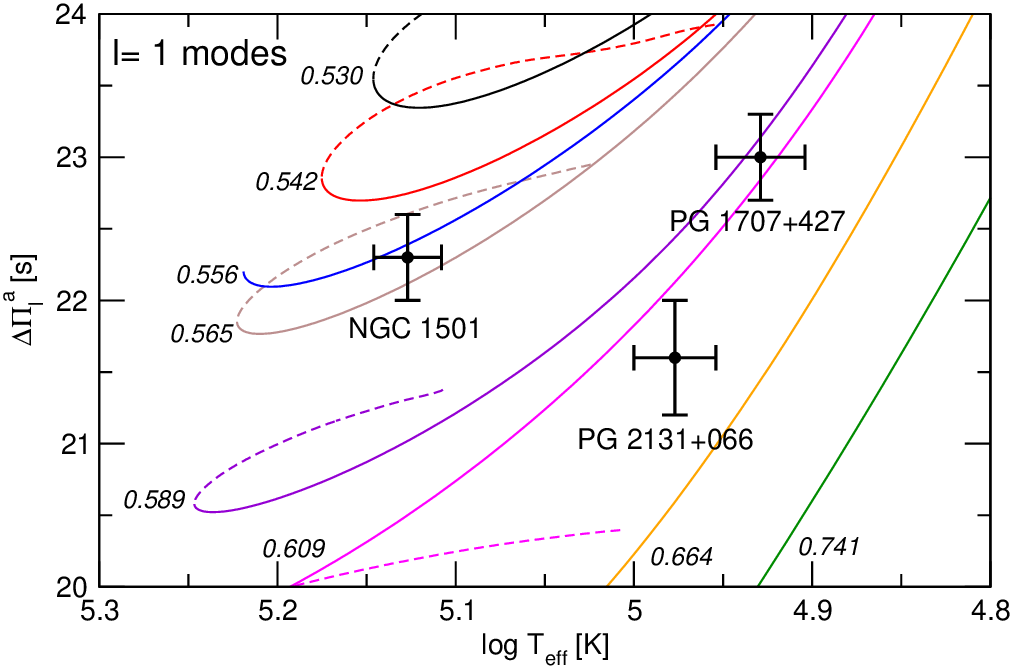}
\caption{The  dipole  asymptotic  period  spacing  in   terms  of  the  
         effective temperature.   Numbers along each  curve denote the
         stellar  mass   (in  solar  units).    Dashed  (solid)  lines
         correspond to evolutionary  stages before (after) the turning
         point  at the  maxima  effective temperature  of each  track.
         Also  shown are  the locations  of \pr,  \pt, and  \ngc.  The
         masses  of  these  stars  as  derived  by  comparing  $\Delta
         \Pi_{\ell=  1}^{\rm a}$ with  $\Delta \Pi^{\rm  O}_{\ell= 1}$
         are  $M_*= 0.627\,  M_{\odot}$, $M_*=  0.597\,  M_{\odot}$, and
         $M_*=  0.571\,  M_{\odot}$, respectively.  Note that  to
         infer the mass of \ngc,  we have considered the stages before
         the evolutionary knee (dashed lines).}
\label{figure1}
\end{figure}

In this section  we constrain the stellar mass of  \pr, \pt, and \ngc\
by  comparing the  asymptotic period  spacing and  the average  of the
computed  period  spacings with  the  {\sl  observed} period  spacing.
These  approaches take  full advantage  of  the fact  that the  period
spacing of PG1159 pulsators depends primarily on the stellar mass, and
the  dependence  on  the  luminosity  and the  He-rich  envelope  mass
fraction is negligible (Kawaler  \& Bradley 1994; C\'orsico \& Althaus
2006).  Most  of the  published asteroseismological studies  on PG1159
stars rely on the asymptotic period spacing to infer the total mass of
GW Vir  pulsators, the  notable exception being  the works by  Reed et
al. (2000)  for \pr, C\'orsico et  al. (2007a) for  \rxj, C\'orsico et
al.  (2007b) for  \pg, and Kawaler \& Bradley  (1994) and C\'orsico et
al.  (2008)  for \pp.  To  assess  the total  mass  of  \ngc\ we  have
considered the  high-luminosity regime of  the evolutionary sequences,
while for  \pr\ and \pt\ we  have focused on the  stages following the
``evolutionary knee''  for the PG 1159 stars,  i.e. the low-luminosity
regime.

\subsection{First method: comparing the observed period 
spacing ($\Delta \Pi^{\rm O}_{\ell}$) with the asymptotic period 
spacing ($\Delta  \Pi_{\ell}^{\rm  a}$)}  
\label{sect-aps}

Fig.  \ref{figure1} displays the  asymptotic period spacing for $\ell=
1$  modes as  a function  of the  effective temperature  for different
stellar masses.   Also shown in this  diagram is the  location of \pr,
with $T_{\rm eff}= 95 \pm 5$ kK (Dreizler \& Heber 1998), and  $\Delta
\Pi^{\rm O}_{\ell= 1}= 21.6 \pm 0.4$ s (Reed et  al. 2000), \pt,  with 
$T_{\rm  eff}= 85  \pm 5$  kK (Dreizler  \& Heber  1998),  and $\Delta
\Pi^{\rm  O}_{\ell= 1}= 23.0  \pm 0.3$  s (Kawaler  et al.  2004), and
\ngc, with $T_{\rm eff}= 134 \pm 5$ kK (Werner \& Herwig 2006), and
$\Delta \Pi^{\rm O}_{\ell= 1}= 22.3 \pm 0.3$ s (Bond et al. 1996). The
asymptotic period spacing is computed as $\Delta \Pi_{\ell}^{\rm a}=
\Pi_0 / \sqrt{\ell(\ell+1)}$, where 

\begin{equation}
\Pi_0= 2 \pi^2 \left[ \int_{r_1}^{r_2} (N/r) dr\right]^{-1}
\end{equation}

\noindent and $N$ is the Brunt-V\"ais\"al\"a frequency (Tassoul et al.  
1990).   From the  comparison  between the  observed $\Delta  \Pi^{\rm
O}_{\ell= 1}$  and $\Delta \Pi_{\ell=  1}^{\rm a}$ we found  a stellar
mass of  $0.627\, M_{\odot}$ for  \pr, $0.597\, M_{\odot}$ for  \pt, and
$0.571 \, M_{\odot}$ for \ngc\ (second column in Table
\ref{tabla-masas}).

\begin{figure}
\centering
\includegraphics[clip,width=230pt]{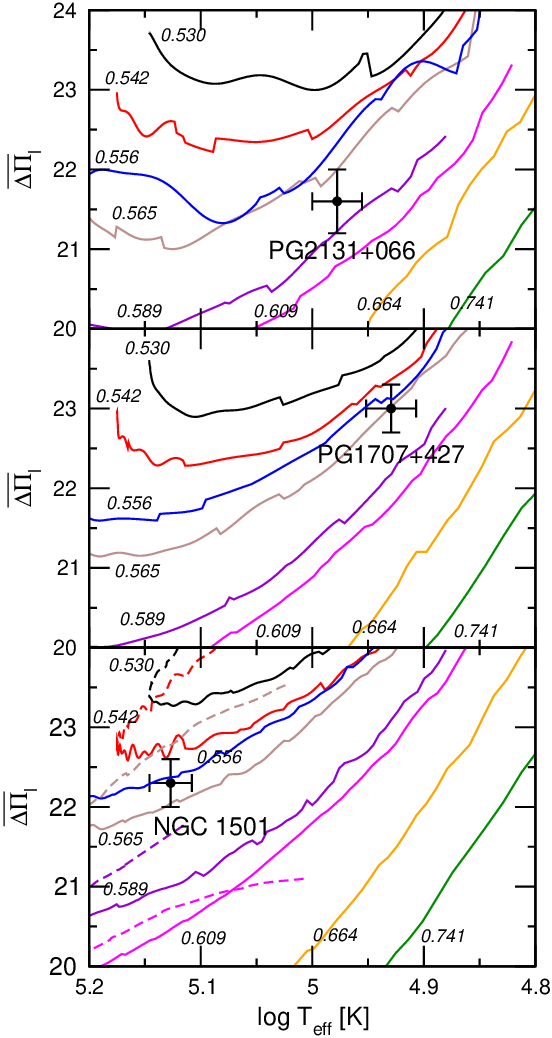}
\caption{Same  as  Fig.  \ref{figure1},  but  for  the  average of the  
         computed period  spacings. For \pr\ and \pt\  only the stages
         after  the  ``evolutionary  knee''  have been  plotted.   The
         masses  of  \pr,  \pt,  and  \ngc\ as  derived  by  comparing
         $\overline{\Delta  \Pi_{\ell=   1}}$  with  $\Delta  \Pi^{\rm
         O}_{\ell=  1}$  are  $M_*= 0.578\,  M_{\odot}$, $M_*=  0.566\,
         M_{\odot}$, and $M_*= 0.576\, M_{\odot}$, respectively.  As 
         in Fig. \ref{figure1},  we  infer   the  mass   of   \ngc\ by 
         considering the stages before the  evolutionary  knee (dashed 
         lines).}
\label{figure2}
\end{figure}

The  method employed  here is  computationally inexpensive  and widely
used because  it does not involve  pulsational calculations.  However,
we must  emphasize that the derivation  of the stellar  mass using the
asymptotic period  spacing may not  be entirely reliable  in pulsating
PG1159  stars  that  pulsate  with  modes  characterized  by  low  and
intermediate  radial  orders  (see  Althaus  et  al.  2007).  This  is
particularly true for \pr\ and \pt.  This shortcoming of the method is
due to that the asymptotic predictions are strictly valid in the limit
of  very   high  radial  order  (long  periods)   and  for  chemically
homogeneous  stellar models,  while PG1159  stars are  supposed  to be
chemically stratified  and characterized by  strong chemical gradients
built up during  the progenitor star life.  A  more realistic approach
to infer the stellar mass of PG1159 stars is presented below.

\subsection{Second method: comparing the observed period 
spacing ($\Delta \Pi^{\rm O}_{\ell}$) with the average of the computed 
period spacings ($\overline{\Delta \Pi_{\ell}}$)}  
\label{sect-psp}

The  average   of  the  computed   period  spacings  is   assessed  as
$\overline{\Delta  \Pi_{\ell}}= (N-1)^{-1}  \sum_{k}  \Delta \Pi_k  $,
where the ``forward'' period spacing is defined as $\Delta \Pi_k=
\Pi_{k+1}-\Pi_k$ ($k$ being the radial order) and $N$ is the number of
computed periods laying in the range of the observed periods. For \pr,
$\Pi_k \in [340,600]$ s, according  to Kawaler et al. (1995); for \pt,
$\Pi_k \in [330,920]$ s, according to Kawaler et al. (2004); and for
\ngc, $\Pi_k \in [1150,2000]$ s, according to Bond et al. (1996). 

 This method is  more reliable for the estimation  of the stellar mass
of PG1159 stars  than that described above because,  provided that the
average  of   the  computed  period  spacings  is   evaluated  at  the
appropriate  range of  periods, the  approach is  appropriate  for the
regimes of short, intermediate and long periods (i.e., $\forall k$) as
well.  When the average of  the computed period spacings is taken over
a  range  of  periods  characterized  by high  $k$  values,  then  the
predictions  of the  present  method  become closer  to  those of  the
asymptotic  period spacing approach.  On the  other hand,  the present
method requires of detailed  period computations, at variance with the
method described in the above  section. In addition, we note that both
methods  for assessing  the  stellar mass  rely  on the  spectroscopic
effective temperature, and the results are unavoidably affected by its
associated uncertainty.

In  Fig.  \ref{figure2} we  show the  run of  average of  the computed
period spacings  ($\ell= 1$) for \pr,  \pt, and \ngc\ in  terms of the
effective temperature  for all  of our PG1159  evolutionary sequences.
The  run of  $\overline{\Delta \Pi_{\ell}}$  depends on  the  range of
periods on which  the average of the computed  period spacing is done.
Note that  the lines shown in  Fig. \ref{figure2} are  very jagged and
jumped. This  is because that,  for a given  star, the average  of the
computed period spacings is evaluated for a fixed period interval, and
not  for a fixed  $k$-interval \footnote{As  the star  evolves towards
higher (lower) effective  temperatures, the periods generally decrease
(increase)  with  time.  At  a  given  $T_{\rm  eff}$, there  are  $N$
computed periods laying in the chosen period interval.  Later when the
model has  evolved enough (heated or  cooled) it is  possible that the
accumulated period drift nearly  matches the period separation between
adjacent modes ($|\Delta k|=  1$).  In these circumstances, the number
of periods laying in the chosen  (fixed) period interval is $N \pm 1$,
and $\overline{\Delta \Pi_{\ell}}$ exhibits a little jump.}.

By adopting the effective temperature of \pr, \pt, and \ngc \ as given
by  spectroscopy we  found  a  stellar mass  of  $0.578 \,  M_{\odot}$,
$0.566\, M_{\odot}$, and $0.576\, M_{\odot}$, respectively.  Our results
are  shown in  the  third column  of  Table \ref{tabla-masas}.   These
values are  $8.5 \%$  (for \pr)  and $5.5 \%$  (for \pt)  smaller than
those  derived through  the  asymptotic period  spacing, showing  once
again that  the asymptotic approach overestimates the  stellar mass of
PG1159 stars that,  like \pr\ and \pt, exhibit  short and intermediate
pulsation periods (see Althaus et  al. 2008a).  On the contrary, there
is a very small discrepancy (in the opposite direction) of $ \sim 0.9
\%$ for the  case of \ngc, showing that in  the long-period regime the
results  for the  stellar mass  obtained using  the  asymptotic period
spacing and the  average of the computed period  spacings nicely agree
each  other.   A similar  situation  is found  in  the  case of  \rxj\
(C\'orsico et al. 2007a).

\subsection{Third method: using an approximate formula}
\label{comparison}

To compare with previous  works, we make an additional estimation
of  the stellar  mass of  \pr, \pt,  and \ngc\  using  the approximate
expression  for the  overall  structure parameter  $\Pi_0$ derived  by
Kawaler \& Bradley (1994):

\begin{equation}
\Pi_0  \approx 15.5 \left(\frac{M_*}{M_{\odot}}\right)^{-1/3}
                    \left(\frac{L_*}{100\,L_{\odot}}\right)^{-0.035}
                    \left(\frac{q_y}{10^{-3}}\right)^{-0.00012}
\label{ecuacion2}
\end{equation}

\noindent where  $q_y$  is the  He-rich envelope  mass fraction.  This 
expression is derived by  considering the dependence of the asymptotic
period spacing on the total mass, stellar luminosity, and thickness of
the He-rich outer envelope for  a large grid of ``quasi evolutionary''
PG1159 models in the  luminosity range $1.6 \la \log(L_*/L_{\odot}) \la
3.0$.  Both the present method and  the method described in Sect. 
\ref{sect-aps} are almost equivalent 
because they are based on the asymptotic period spacing.

Due  to the  very  weak dependence  of  $\Pi_0$  on $q_y$,  we
arbitrarily fix  it to a value  of $10^{-2}$. Since  the luminosity is
not known at the outset, we can compute it as $L_*= 4 \pi \sigma R_*^2
T_{\rm eff}^4$, where  $R_*^2= G M_*/g$. We use the  values of $g$ and
$T_{\rm eff}$ inferred through spectroscopy.  Assuming that $\Pi_0$ is
known  from  the  observed  period  spacing  ($\Pi_0  \sim  \sqrt{\ell
(\ell+1)}\ \Delta \Pi^{\rm O}_{\ell}$), we obtain an estimation of the
stellar mass from Eq. (\ref{ecuacion2}).  Our results are shown in the
fourth column  of Table  \ref{tabla-masas}.   For  \pr\  and \pt\  the
stellar masses  obtained in this way  are in very  good agreement with
our values  derived from the  asymptotic period spacing (first  row in
Table \ref{tabla-masas}).   This is not an  unexpected result because,
as  mentioned, the  expression of  Kawaler \&  Bradley (1994)  is also
based on the asymptotic  period spacing.  The slight differences found
(below $\approx 2.5  \%$) could be due to  differences in the modeling
of PG 1159 stars. For \ngc, instead, there is a substantial difference
($\approx 7 \%$) between the  prediction of this formula and our value
inferred from the asymptotic period spacing.  This could be due to the
inadequacy  of  the formula  of  Kawaler  \&  Bradley (1994)  for  the
high-luminosity regime characterizing the evolutionary status of \ngc. 
 
\section{Constraints from the individual observed periods}
\label{fitting}

 In this  approach we seek pulsation  models that best  match the {\sl
individual} pulsation  periods of \pr,  \pt, and \ngc.  For  the three
stars, we assume that all of the observed periods correspond to $\ell=
1$  modes  because the  observed  period  spacings  and the  frequency
splittings by rotation  are consistent with $\ell= 1$  (Kawaler et al.
1995; Kawaler et al. 2004; Bond et al. 1996).  To measure the goodness
of the  match between  the theoretical pulsation  periods ($\Pi_k^{\rm
T}$) and the observed  individual periods ($\Pi_i^{\rm O}$), we follow
the same  $\chi^2$ procedure  as in our  previous works ---  see, e.g,
C\'orsico  et   al.  (2007a)    and  Townsley  et   al.  (2004).
Specifically, we employ the quality function defined as

\begin{equation}
\chi^2(M_*,T_{\rm eff})= \frac{1}{n}\sum_{i=1}^{n}   
\min\left[(\Pi_i^{\rm O}-\Pi_k^{\rm T})^2\right]
\end{equation}

\noindent where $n$  is the number of  observed  periods. 
The  observed periods are shown in the first
column of Tables  \ref{table1} and \ref{table2} for \pr\  and \pt, and
in Table 7 of Bond et al. (1996) for \ngc. 

Next, we briefly explain the procedure we follow to found a model
representative of a target star. For a given model (characterized by a
given  $T_{\rm eff}$)  corresponding  to an  evolutionary sequence  of
stellar mass $M_*$, we consider the first observed period of the list,
namely $\Pi_1^{\rm O}$, and compute the successive squared differences
$(\Pi_1^{\rm O}-\Pi_k^{\rm  T})^2$, where the radial  order 
$k$ varies from $1$ to a given maximum value, $k_{\rm max}$, which 
corresponds to
a theoretical  period far longer  that the maximum period  observed in
the star.  Next,  we retain the minor squared  difference, and then we
repeat  the procedure but  this time  considering the  second observed
period,  namely $\Pi_2^{\rm  O}$. After  the minor  squared difference
associated with  this observed period  is stored, we proceed  with the
next  observed  period, and  so  until  the  minor squared  difference
associated  with  the  last   observed  period  
$(\Pi_n^{\rm  O})$  is
stored. The next step is to calculate the sum of these differences and
then obtain the value of $\chi^2(M_*,T_{\rm eff})$ for the model under
consideration.   It is worth 
mentioning that
with this algorithm, the value of $\chi^2(M_*,T_{\rm eff})$ does not 
depend on the particular order in which the observed periods are fitted.
The complete  algorithm is  
repeated for  all  of the models of the sequence, and then, 
a curve of $\chi^2(M_*,T_{\rm eff})$
versus  $T_{\rm eff}$  is obtained  for the  complete  sequence.  This
procedure is  carried out for all  of our sequences. For  each star of
interest, the PG 1159 model that shows the lowest value of $\chi^2$ is
adopted  as  the  ``best-fit  model''.

\begin{figure}
\centering
\includegraphics[clip,width=230pt]{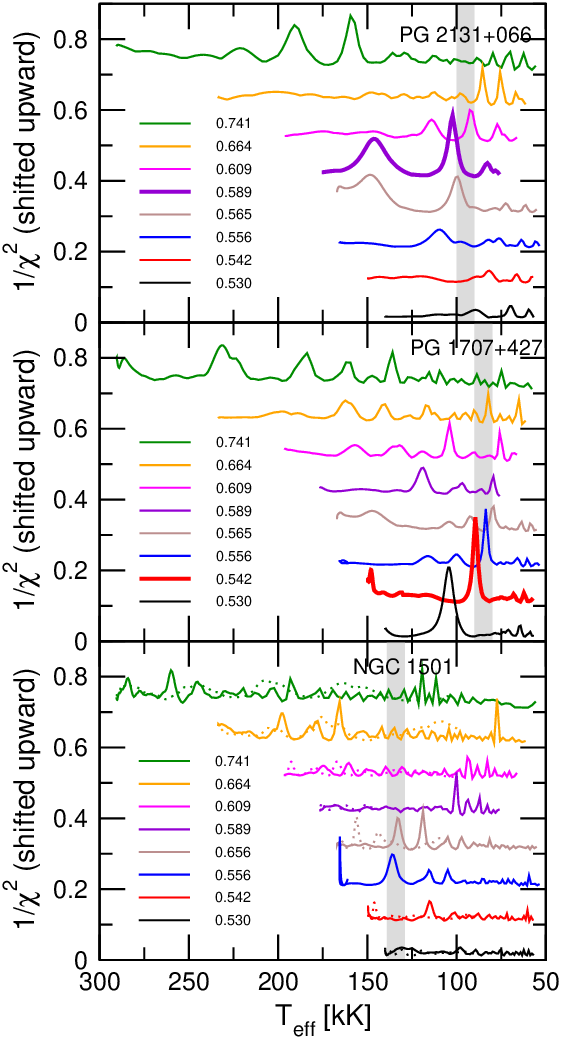}
\caption{The inverse of  the  quality  function  of  the period fit in 
         terms of  the effective  temperature (see text  for details).
         The vertical  grey strip indicates  the spectroscopic $T_{\rm
         eff}$   and  its   uncertainties.   The   curves   have  been
         arbitrarily shifted upward (with  a step of 0.1). Upper panel
         corresponds  to \pr,  middle  panel corresponds  to \pt,  and
         lower panel  corresponds to \ngc.   For  \ngc, the stages
         before  the  evolutionary  knee  are  displayed  with  dotted
         lines.}
\label{chi2} 
\end{figure}

\subsection{The search for the best-fit models}
\label{searching}

We evaluate the function $\chi^2=\chi^2(M_*, T_{\rm eff})$ for stellar
masses  of $0.530,  0.542,  0.556, 0.565,  0.589,  0.609, 0.664$,  and
$0.741 M_{\odot}$.   For the effective temperature we  employed a much
finer   grid  ($\Delta   T_{\rm   eff}=  10-30$   K).   The   quantity
$(\chi^2)^{-1}$ in  terms of  the effective temperature  for different
stellar masses  is shown in  Fig.  \ref{chi2} for \pr\  (upper panel),
\pt\  (middle  panel), and  \ngc\  (lower  panel),  together with  the
corresponding spectroscopic effective  temperatures.  We prefer to
show  in our plots  the quantity  $(\chi^2)^{-1}$ instead  $\chi^2$ in
order to  emphasize the location  of models providing  good agreements
between observed and theoretical periods.   As mentioned,  
the goodness of the match between the observed and theoretical 
periods is reflected by the value of $\chi^2$. The lower the value of 
$\chi^2$, the better the period match.  We will consider ---admittedly 
somewhat arbitrarily--- that a  peak in the quality function with 
$\chi^2 \lesssim 5$ (that is, $(\chi^2)^{-1} \gtrsim 0.2$)
is a good match between the  theoretical and the observed periods.

For \pr\  we find  one strong maximum  of $(\chi^2)^{-1}$ for  a model
with $M_*= 0.589 \, M_{\odot}$ and $T_{\rm eff}\approx 102$ kK.  Such a
pronounced  maximum in the  inverse of  $\chi^2$ implies  an excellent
agreement between  the theoretical and observed  periods. Another much
less pronounced maxima, albeit at effective temperatures closer to the
spectroscopic estimation  for \pr, are encountered for  $M_*= 0.609 \,
M_{\odot}$ at $T_{\rm eff}\approx  92.4$ kK and $M_*= 0.565\, M_{\odot}$
at  $T_{\rm  eff}\approx 99.8$  kK.   However,  because the  agreement
between  observed  and  theoretical   periods  for  these  models  are
substantially poorer than for the one with $M_*= 0.589\, M_{\odot}$, we
adopt  this last model  as the  best-fit asteroseismological  model. A
detailed comparison  of the observed $\ell=  1, m= 0$  periods in \pr\
with  the theoretical  periods of  the best-fit  model is  provided in
Table  \ref{table1}.  The  high  quality   of  our  period   fit  is
quantitatively  reflected  by  the  average  of  the  absolute  period
differences

\begin{equation}
\overline{\delta \Pi_i}= \frac{1}{n}\sum_{i=1}^n \left|\delta\Pi_i\right| 
\end{equation}

\noindent where $\delta \Pi_i= \Pi_i^{\rm O} -\Pi_k^{\rm T}$ and by the 
root-mean-square residual

\begin{equation}
\sigma_{_{\delta \Pi_i}}= \sqrt{\frac{(\sum_{i=1}^n |\delta \Pi_i|^2)}{n}}. 
\end{equation}

\noindent  Note that $\sigma_{_{\delta \Pi_i}}$ is simply
the function $\sqrt{\chi^2}$ evaluated at the best-fit model. For 
the best-fit model of \pr\  we  obtain  $\overline{\delta \Pi_i}= 
1.57$ s  and $\sigma_{_{\delta\Pi_i}}= 2.32$ s, which  are indeed very
small.

For \pt\  we find  one strong maximum  of $(\chi^2)^{-1}$ for  a model
with $M_*=  0.542 \,  M_{\odot}$ and $T_{\rm  eff}\approx 89.5$  kK, an
effective temperature compatible with the spectroscopic determination.
Another somewhat  less pronounced  maximum is found  for a  model with
$M_*= 0.556  \, M_{\odot}$ and  $T_{\rm eff}\approx 83.6$  kK.  Despite
the fact  that this model has  an effective temperature  very close to
the  spectroscopic  one,  we  choose  the  model  with  $M_*=  0.542\,
M_{\odot}$ as  the best-fit  model for \pt,  because the period  fit is
characterized  by  a  better  quality.   Table  \ref{table2}  shows  a
comparison between  observed $\ell= 1,  m= 0$ and computed  periods of
the best-fit  model. We found  in this case  $\overline{\delta \Pi_i}=
1.75$ s and $\sigma_{_{\delta \Pi_i}}= 1.99$ s.

The situation for  \ngc\ is markedly different than  for \pr\ and \pt.
Indeed, as  clearly shown in the  lower panel of  Fig. \ref{chi2}, for
this star the $(\chi^2)^{-1}$  function exhibits numerous local maxima
at several  values of the  effective temperature and the  stellar mass
($M_*/M_{\odot}=  0.556$, 0.565,  0.589,  0.664, and  0.741) that  have
roughly the same amplitudes,  making virtually impossible to isolate a
clear and  unambiguous seismological solution. Thus, for  \ngc\ we are
unable  to find  a best-fit  seismological model.   This could  be, in
part,  due to  the fact  that  the periods  detected in  \ngc\ can  be
associated to  eigenmodes with radial orders $k$  quite different from
each other  (with a mean spacing  of $\overline{\Delta k}  \ga 6$), in
such a way  that it is easy to find  numerous models (characterized by
strongly different $T_{\rm  eff}$ and $M_*$) that reproduce  to a some
extent the  observed period spectrum of  \ngc.  It could  also be that
the  impossibility to  find a  best fit  model would  be  reflecting a
different  evolutionary history for  \ngc\ than  that assumed  in this
work for our PG1159 sequences.

\begin{table}
\centering
\caption{Observed and  theoretical ($\ell=1$)  periods of the best-fit 
         model for PG  2131+066 ($M_*= 0.589M\,_{\odot}$, $T_{\rm eff}=
         102\,179$  K,  $\log(L_*/L_{\odot})=  1.57$). Periods  are  in
         seconds and rates of period change (theoretical) are in units
         of  $10^{-12}$  s/s.  The  observed  periods  are taken  from
         Kawaler et al. (1995).}
\begin{tabular}{ccccc}
\hline
\hline
\noalign{\smallskip}
$\Pi_i^{\rm O}$ & $\Pi_k^{\rm T}$ & $k$ & $\dot{\Pi}^{\rm T}_k$ & 
unstable\\
\noalign{\smallskip}
\hline
\noalign{\smallskip}
$ 341.45 $ & $341.88  $ & 14 & $2.04 $ &  yes     \\
---        & $363.66  $ & 15 & $3.90 $ &  yes     \\   
$ 384.27 $ & $383.94  $ & 16 & $2.60 $ &  yes     \\
$ 403.93 $ & $403.30  $ & 17 & $3.30 $ &  yes     \\
$ 426.36 $ & $426.48  $ & 18 & $4.14 $ &  yes     \\
$ 450.28 $ & $445.31  $ & 19 & $2.89 $ &  yes     \\
$ 462.39 $ & $465.91  $ & 20 & $4.53 $ &  yes     \\
---        & $488.81  $ & 21 & $4.11 $ &  no      \\   
$ 507.91 $ & $507.91  $ & 22 & $3.79 $ &  no      \\
\hline
\hline
\end{tabular}
\label{table1}
\end{table}

The fourth  column in Tables  \ref{table1} and \ref{table2}  shows the
rates of period change associated with the fitted modes for \pr\ and
\pt, respectively.  Our calculations predict that all of the pulsation
periods {\sl increase} with time ($\dot{\Pi} > 0$), in accordance with
the decrease of  the Brunt-V\"ais\"al\"a frequency in the  core of the
models induced by cooling. At the effective temperatures of \pr\ and
\pt, cooling has the largest effect on $\dot{\Pi}$, while
gravitational contraction, which should  result in a {\sl decrease} of
periods  with  time, becomes  negligible  and  no  longer affects  the
pulsation periods, except  for the case of modes  {\sl trapped} in the
envelope.  Until  now, the only secure measurements  of $\dot{\Pi}$ in
pre-white dwarf stars are those of \pp, the prototype of the class, by
Costa et al.  (1999) and more  recently by Costa \& Kepler (2008).  In
this last  paper the authors obtained  a mix of  positive and negative
values  of $\dot{\Pi}$,  indicating that  for that  star gravitational
contraction is still important. In principle, the needed time interval
that  the  observational  data  should  cover  in  order  to  reach  a
measurement of a $\dot \Pi$ in PG  1159 stars like \pr\ and \pt\ is of
about ten  years. Unfortunately, no  future observations in  the short
term that could  allow a determination of $\dot{\Pi}$  for these stars
and thus to check the predictions of our models are foreseen.

\begin{table}
\centering
\caption{Same as Table \ref{table1}, but  for  the best-fit  model for 
         PG 1707+427  ($M_*=0.542\, M_{\odot}$, $T_{\rm  eff}= 89\,504$
         K,  $\log(L_*/L_{\odot})= 1.40$).   The  observed periods  are
         taken from Kawaler et al. (2004).}
\begin{tabular}{ccccc}
\hline
\hline
\noalign{\smallskip}
$\Pi_i^{\rm O}$ & $\Pi_k^{\rm T}$ & $k$ & $\dot{\Pi}^{\rm T}_k$ & 
unstable\\
\noalign{\smallskip}
\hline
\noalign{\smallskip}
$ 334.62 $ & $ 333.26 $ & 12 & $2.13 $ &  yes     \\
  ---      & $ 355.73 $ & 13 & $1.19 $ &  yes     \\  
  ---      & $ 378.53 $ & 14 & $2.17 $ &  yes     \\  
  ---      & $ 401.64 $ & 15 & $1.51 $ &  yes     \\  
  ---      & $ 423.68 $ & 16 & $1.84 $ &  yes     \\  
$ 448.07 $ & $ 447.31 $ & 17 & $2.46 $ &  yes     \\
  ---      & $ 469.26 $ & 18 & $1.49 $ &  yes     \\  
$ 494.39 $ & $ 492.56 $ & 19 & $2.76 $ &  yes     \\
  ---      & $ 516.50 $ & 20 & $2.61 $ &  yes     \\  
$ 536.41 $ & $ 537.67 $ & 21 & $1.66 $ &  yes     \\
  ---      & $ 562.32 $ & 22 & $3.30 $ &  yes     \\  
  ---      & $ 584.78 $ & 23 & $2.37 $ &  yes     \\  
  ---      & $ 606.62 $ & 24 & $2.73 $ &  yes     \\  
  ---      & $ 632.11 $ & 25 & $3.25 $ &  yes     \\  
  ---      & $ 653.15 $ & 26 & $2.99 $ &  yes     \\  
$ 677.89 $ & $ 677.38 $ & 27 & $3.43 $ &  yes     \\
  ---      & $ 701.20 $ & 28 & $3.03 $ &  yes     \\  
$ 726.02 $ & $ 722.82 $ & 29 & $3.95 $ &  no      \\
$ 745.78 $ & $ 747.58 $ & 30 & $3.42 $ &  no      \\
  ---      & $ 770.93 $ & 31 & $3.71 $ &  no      \\  
  ---      & $ 793.13 $ & 32 & $4.45 $ &  no      \\  
  ---      & $ 817.86 $ & 33 & $3.61 $ &  no      \\  
  ---      & $ 841.70 $ & 34 & $4.64 $ &  no      \\  
  ---      & $ 863.34 $ & 35 & $3.91 $ &  no      \\  
  ---      & $ 888.59 $ & 36 & $5.03 $ &  no      \\  
$ 909.05 $ & $ 912.34 $ & 37 & $4.21 $ &  no      \\
\hline
\hline
\end{tabular}
\label{table2}
\end{table}

Finally, the last column in Tables \ref{table1} and \ref{table2} gives
information about the  pulsational stability/instability nature of the
modes associated with  the periods fitted to the  observed ones.  Full
nonadiabatic  calculations employing the  pulsation code  described in
C\'orsico et  al. (2006) indicate that,  for the case of  \pr, all the
fitted modes except one (that with period 507.9 s) are predicted to be
unstable.  For the case of \pt, our nonadiabatic computations are able
to  explain the  existence  of  periodicities in  the  range $330  \la
\Pi^{\rm O}  \la 680$ s only,  while they fail  to predict pulsational
instability for the observed modes with periods at 726, 746, and 909 s.

\subsection{Characteristics of the best-fit models for \pr\ and \pt}
\label{char}

The main  features of  our best-fit model  for \pr\ are  summarized in
Table \ref{table3}, where  we also provide the parameters  of the star
extracted  from other  published  studies. Specifically, the   
second   column corresponds to spectroscopic  results  from Werner \& Herwig
(2006), whereas  the third and fourth  columns present results
from the pulsation studies of  Kawaler et al. (1995) and Reed
et al. (2000), and from the asteroseismological model of this
work,   respectively. The number in parenthesis is the 
spectroscopic estimation of the stellar mass employing 
the evolutionary tracks of Miller Bertolami \& Althaus (2006).
In  the present  work, errors  in
$T_{\rm eff}$  and $\log(L_*/L_{\odot})$ are estimated  from the width
of the maximum in the function $\chi^2$ with respect $T_{\rm eff}$ and
$\log(L_*/L_{\odot})$,  respectively.  The  error in  the  stellar mass
comes from  the grid resolution  in $M_*$. Errors  in the rest  of the
quantities are  derived from these values.   The effective temperature
of our best-fit model ($T_{\rm  eff}= 102\, 180$ K) is somewhat higher
than ---but still compatible with--- the spectroscopic value ($T_{\rm
eff}= 95\, 000  \pm 5\,000$ K).  On the other hand,  the total mass of
the best-fit model ($M_*= 0.589 \, M_{\odot}$) is in agreement with the
value derived  from the  average of the computed period spacings ($M_*
\sim  0.578  \, M_{\odot}$), but  at  odds ($\sim 6 \%$  smaller)  with 
that  inferred  from  the  asymptotic period  spacing  ($M_*=  0.627\,
M_{\odot}$) (see  Table \ref{tabla-masas}).   Also, the $M_*$  value of
our best-fit model is substantially larger than the spectroscopic mass
of $0.55 \, M_{\odot}$ derived by Miller Bertolami \& Althaus (2006),
but very similar to $0.58 \, M_{\odot}$ 
according to Werner \& Herwig (2006).    A    
discrepancy   between   the asteroseismological and the spectroscopic 
values of $M_*$ is generally
encountered among  PG1159 pulsators ---  see C\'orsico et  al.  (2006,
2007ab).  Until now, the asteroseismological mass of \pr\ has been about
$11  \%$  larger  ($\Delta  M_*  \approx 0.06\,  M_{\odot}$)  than  the
spectroscopic  mass  if  we  consider  the early  estimation  for  the
seismological  mass quoted by  Reed et  al. (2000) and the derivation 
of Miller Bertolami \& Althaus (2006) for the spectroscopic 
mass\footnote{We elect the value of the spectroscopic mass 
of \pr\  inferred by 
Miller Bertolami \& Althaus (2006) for this comparison 
because they use the same post-born again PG1159 evolutionary models 
we employ here in the determination of the asteroseismological mass, 
and because the spectroscopic masses quoted by Werner \& Herwig 
(2006) are based on old post-AGB tracks.}.   
In light  of the
best-fit  model derived in  this paper,  this discrepancy  is slightly
reduced  to  about $7  \%$  ($\Delta  M_*  \approx 0.04\,  M_{\odot}$).
Finally, our  best-fit model  for \pr\ is  somewhat more  luminous and
less compact  than what  is suggested  by the results  of Reed  et al.
(2000).  

\begin{table}
\begin{center}
\caption{ The  main   characteristics   of  \pr.}
\begin{tabular}{lccc}
\hline
\hline
\noalign{\smallskip}
Quantity                     & Spectroscopy                  & Seismology              & This work                 \\
\hline
\noalign{\smallskip}
$T_{\rm eff}$ [kK]           & $95 \pm 5$                    &        ---              & $102.18_{-2.8}^{+3.0}$    \\
\noalign{\smallskip}
$M_*$ [$M_{\odot}$]           & $0.58\pm 0.1$                 & $0.608 \pm 0.01$        & $0.589_{-0.024}^{+0.020}$ \\ 
\noalign{\smallskip}
                             & ($0.55\pm 0.1$)               &                         &                           \\   
\noalign{\smallskip}
$\log g$ [cm/s$^2$]          & $7.5 \pm 0.5$                 &        ---              & $7.63_{-0.14}^{+0.12} $   \\ 
\noalign{\smallskip}
$\log (L_*/L_{\odot})$        & ---                           & $1.41 \pm 0.5$          & $1.57_{-0.06}^{+0.07}$    \\  
\noalign{\smallskip}
$\log(R_*/R_{\odot})$         & ---                           & $-1.73$                 & $-1.71_{-0.05}^{+0.06}$   \\  
\noalign{\smallskip}
$M_{\rm env}$ [$M_{\odot}$]   & ---                           & $3.8 \times 10^{-3}$    & $0.016$                   \\  
\noalign{\smallskip}
$M_{\rm V}$ [mag]            & ---                           & $7.69^{+0.25}_{-0.18}$  & $6.825_{-0.675}^{+0.655}$ \\
\noalign{\smallskip}
$M_{\rm bol}$ [mag]          & ---                           &        ---              & $0.825_{-0.175}^{+0.155}$ \\
\noalign{\smallskip}
$A_{\rm  V}$ [mag]           & ---                           &        ---              & $0.18$                    \\ 
\noalign{\smallskip}
$d$  [pc]                    & ---                           & $668_{-83}^{+78}$       & $830_{-224}^{+300}$       \\ 
\noalign{\smallskip}
$\pi$ [mas]                  & ---                           & $1.50\pm 0.2$           & $1.2_{-0.3}^{+0.4}$       \\ 
\noalign{\smallskip}
\hline
\hline
\end{tabular}
\label{table3}
\end{center}
\end{table}

The main properties of our best-fit  model for \pt\ are shown in Table
\ref{table4}.  The   
second   column corresponds to spectroscopic  results  from Werner \& Herwig
(2006), whereas  the third and fourth  columns present results
from the pulsation study of  Kawaler et al. (2004) and from the 
asteroseismological model of this work,   respectively. 
As for the case of \pr, the effective temperature of our
best-fit  model for  \pt\ is  slightly larger  than  the spectroscopic
measurement, but even in good agreement with it. Regarding the stellar
mass,  our best-fit  model has  $M_*= 0.542\,  M_{\odot}$, which  is in
agreement  with the  value derived  from the  average of  the computed
period spacings  ($M_* \sim 0.566\,  M_{\odot}$), but at odds  ($\sim 9
\%$  lower) with  that  inferred from  the  asymptotic period  spacing
($M_*= 0.597 \, M_{\odot}$) (see Table \ref{tabla-masas}). On the other
hand,  we note  that  $M_*$ for  the  best-fit model  is in  excellent
agreement  with  the  spectroscopic  derivation of Miller Bertolami \& 
Althaus (2006) ($0.542  \,  M_{\odot}$ versus $0.53  \, M_{\odot}$), but 
is substantially lower than the 
spectroscopic value quoted by Werner \& Herwig (2006) 
($0.59  \,  M_{\odot}$).  Until now,  the asteroseismological mass
of \pt\ has  been more than $7 \%$ larger  ($\Delta M_* \approx 0.04\,
M_{\odot}$)  than   the  spectroscopic  mass   if  we  adopt   for  the
seismological mass the value found by Kawaler et al. (2004) and  
the derivation of Miller Bertolami \& Althaus (2006) for the 
spectroscopic mass.  In light
of our best-fit  model, this discrepancy is strongly  reduced to about
$2  \%$  ($\Delta  M_*  \approx  0.012 \,  M_{\odot}$).   Finally,  our
best-fit  model  for \pt\  is  slightly  more  luminous than  what  is
suggested by Kawaler  et al. (2004).

\begin{table}
\begin{center}
\caption{Same  as  Table  \ref{table3}, but for \pt.}
\begin{tabular}{lccc}
\hline
\hline
\noalign{\smallskip}
Quantity                     & Spectroscopy                     & Seismology              & This work                 \\
\hline
\noalign{\smallskip}
$T_{\rm eff}$ [kK]           & $85 \pm 4.5$                     &        ---              & $89.5_{-1.8}^{+1.7}$      \\
\noalign{\smallskip}
$M_*$ [$M_{\odot}$]          & $0.59\pm 0.1$                    & $0.57 \pm 0.02$         & $0.542_{-0.012}^{+0.014}$ \\ 
\noalign{\smallskip}
                             & ($0.53\pm 0.1$)                  &                         &                           \\ 
\noalign{\smallskip}
$\log g$ [cm/s$^2$]          & $7.5 \pm 0.3$                    &        ---              & $7.53_{-0.08}^{+0.09}$    \\ 
\noalign{\smallskip}
$\log (L_*/L_{\odot})$       & ---                              & $1.36$                  & $1.40\pm 0.04$            \\  
\noalign{\smallskip}
$\log(R_*/R_{\odot})$        & ---                              &      ---                & $-1.68\pm 0.04$           \\  
\noalign{\smallskip}
$M_{\rm env}$ [$M_{\odot}$]  & ---                              & ---                     & $0.025$                   \\  
\noalign{\smallskip}
$M_{\rm V}$ [mag]            & ---                              &        ---              & $7.25\pm 0.6$             \\
\noalign{\smallskip}
$M_{\rm bol}$ [mag]          & ---                              &        ---              & $1.25\pm 0.1$             \\
\noalign{\smallskip}
$A_{\rm  V}$ [mag]           & ---                              &        ---              & $0.12$                    \\ 
\noalign{\smallskip}
$d$  [pc]                    & ---                              & ---                     & $730_{-175}^{+230}$       \\ 
\noalign{\smallskip}
$\pi$ [mas]                  & ---                              & ---                     & $1.4\pm0.4$               \\ 
\noalign{\smallskip}
\hline
\hline
\end{tabular}
\label{table4}
\end{center}
\end{table}

\subsection{The asteroseismological distance and parallax of \pr\ and \pt}
\label{distance}

 As in our previous works ---  see, e.g., C\'orsico et al. (2007a) ---
we employ the  luminosity of our best-fit models  to infer the seismic
distance to \pr\  and \pt. Following Kawaler et  al.  (1995), we adopt
${\rm BC}=  -6.0 \pm 0.5$  for both stars  (Werner et al.   1991).  We
account  for  the  interstellar  absorption, $A_{\rm  V}$,  using  the
interstellar extinction model of Chen  et al.  (1998).  With all these
ingredients the  seismic distance, $d$,  can be easily  computed using
the  apparent magnitudes,  which are  $m_{\rm v}=  16.6$ for  \pr\ and
$m_{\rm v}= 16.7$  for \pt\ (Bond et al. 1984).   We obtain a distance
$d \sim  830$ pc and  an interstellar extinction $A_{\rm  V}\sim 0.18$
for \pr\, and $d \sim 730$ pc and $A_{\rm V} \sim 0.12$ for \pt.

Our  estimation of the  distance to \pr\  is $\sim 15  \%$ larger
than that  derived by  CA06 ($d  \sim 716$ pc).   This is  because our
asteroseismological model is somewhat  more luminous than that of CA06
($\log(L_*/L_{\odot})= 1.57$  versus $\log(L_*/L_{\odot})= 1.37$).  On
the other hand,  our distance is $20-25 \%$  larger than that obtained
by Reed  et al. (2000)  ($d \sim  668$ pc) on  the basis of  their own
asteroseismological analysis. This difference can be understood on the
basis   that    Reed   et   al.   (2000)   uses    a   luminosity   of
$\log(L_*/L_{\odot})  \sim  1.4$,  somewhat  lower than  that  of  our
best-fit  model for \pr,  of $\log(L_*/L_{\odot})  \sim 1.6$.   On the
other  hand, our asteroseismological  distance for  \pr\ is  about 1.2
times longer than  that quoted by Reed et al. (2000)  of $\sim 680$ pc
obtained on  the basis of spectrum fitting,  although both estimations
are compatible at the $1 \sigma$ level.

For \pt,  our asteroseismological distance  is in agreement  with that
quoted  by CA06,  of $\sim  697$  pc. Werner  et al.  (1991) obtain  a
distance  to \pt\  of $\sim  1300$ pc,  substantially larger  than our
estimation,  but still  within the  quoted error  bars.  The different
value of Werner et al. (1991) is  due to that they use a luminosity of
$\log L_*/L_{\odot}=  2.15$, substantially higher  than the luminosity
of our best fit model ($\log L_*/L_{\odot}= 1.4$).

\section{Comparison with the results of CA06}
\label{comparison}

\begin{figure}
\centering
\includegraphics[clip,width=250pt]{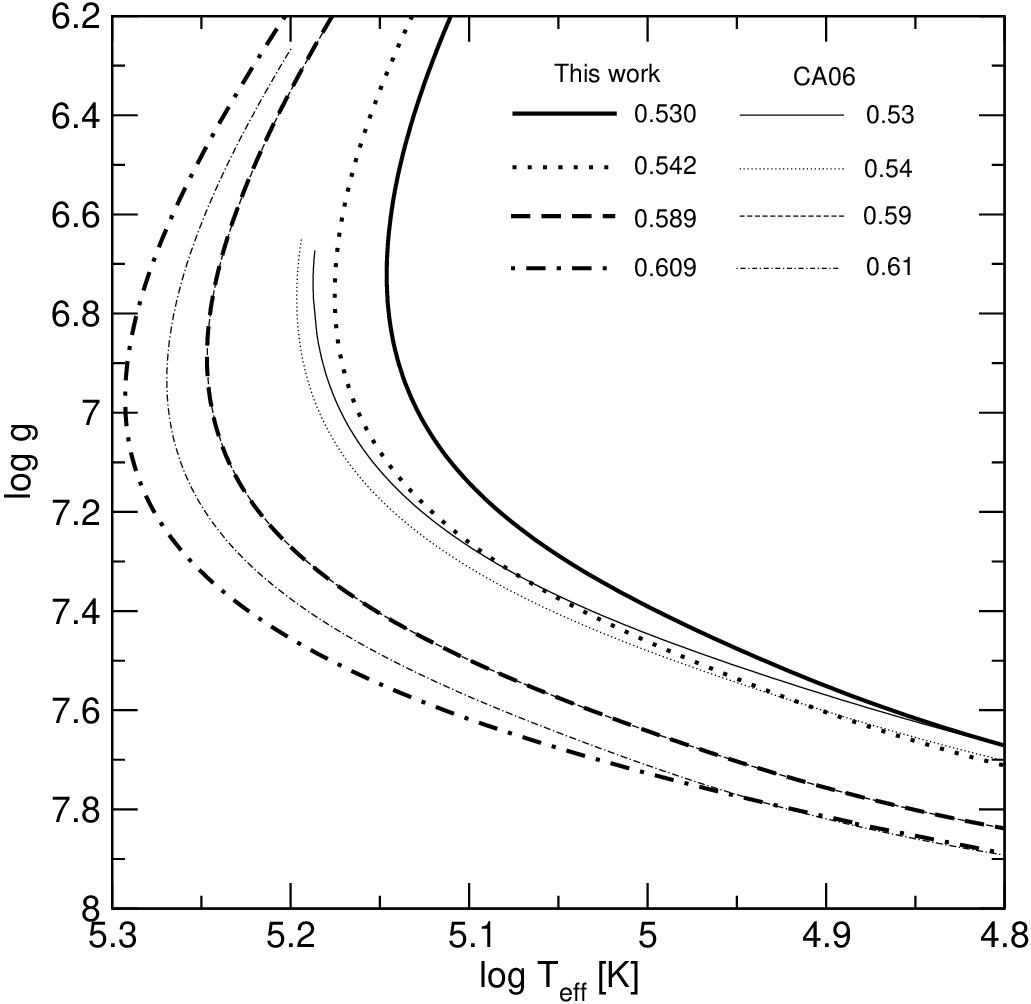}
\caption{Comparison between some PG1159 evolutionary tracks employed in 
the present work (thick lines) and those of CA06 (thin lines).
The values are in solar masses. Note that the track 
corresponding to the sequence of $\approx 0.59 M_{\odot}$ is the same 
for the two sets of computations.}
\label{fig04} 
\end{figure}

Following the recommendations of an anonymous referee, we include
in this  section a detailed  comparison between the PG1159  models and
the asteroseismological results of the  present paper and those of the
previous study by CA06. These authors performed an asteroseismological
analysis of four GW Vir stars (\pg, \pp, \pr, and \pt) on the basis of
a set  of twelve PG1159 evolutionary sequences  with different stellar
masses  ($M_*=   0.53,  0.54,0.55,\cdots,0.62,0.63,  0.64  M_{\odot}$)
artificially derived  from the  full evolutionary sequence  of $0.5895
M_{\odot}$ computed by Althaus et  al. (2005). That sequence is one of
the  sequences  we  use  in  the  present  paper.   Specifically,  the
sequences  of  CA06 were  constructed  using  LPCODE by  appropriately
scaling the stellar mass of  the $0.589 M_{\odot}$ sequence before the
models  reach the  low-luminosity, high-gravity  stage of  the  GW Vir
domain.   Although this  procedure  leads to  a  series of  unphysical
stellar  models  for  which   the  helium-burning  luminosity  is  not
consistent with the  thermo-mechanical structure, the transitory stage
vanishes shortly before the  star reaches the evolutionary ``knee'' in
the HR  diagram (see Fig. 2  of CA06). As a  consequence, those PG1159
models were  not suitable for the  high-luminosity, low-gravity regime
corresponding, for instance, to \rxj, \ngc, K 1-16, HE 1429-1209, etc.

Because  the sequences  of  CA06 with  different  stellar masses  were
created starting  from a single sequence with  $M_*= 0.589 M_{\odot}$,
the central abundances of C and O and the spatial extension of the C-O
core are not completely consistent with the value of the stellar mass,
except in  the case  of the models  of the $0.589  M_{\odot}$ sequence
itself.  For instance, a model  of CA06 with $M_*= 0.64 M_{\odot}$ has
a C-O core that is somewhat  smaller and the central abundance of O is
substantially  higher than  what  would be  expected  if the  complete
evolution of the progenitor star were performed, as it is the case for
the models employed in the  present paper. Fig.  2 of Miller Bertolami
\& Althaus (2006) clearly illustrate that, when the complete evolution
of  the  PG1159 progenitor  stars  is  taken  into account,  different
stellar masses  are associated with different central  abundances of C
and O,  and different  sizes of the  C-O core. Specifically,  the more
massive the models, the lower  (higher) the central abundance of O (C)
and the larger  the C-O core. In summary, for a  given value of $M_*$,
and
\emph{for stages after the evolutionary knee}, the only 
structural/physical difference  between the PG1159  models employed in
the present work and  those of CA06 is related to the  size of the C-O
core and the central abundances of O and C.

In Fig.  \ref{fig04} we compare some evolutionary tracks ($M_*= 0.530,
0.542, 0.589,  0.609 M_{\odot}$) of  the PG1159 sequences  employed in
the present  work with the  corresponding evolutionary tracks  of CA06
($M_*=  0.53, 0.54, 0.59,  0.61 M_{\odot}$).  A careful  inspection of
this figure  reveals that  both sets of  tracks generally  differ, but
when the models reach the  beginning of their white dwarf stage ($\log
T_{\rm eff} \lesssim 5$), they  turn be in very close agreement.  This
agreement is  reached earlier  in the case  of sequences  with stellar
masses close to $M_*= 0.589 M_{\odot}$, the value of the sequence from
which the  remainder sequences of CA06  were generated.  Interestingly
enough, the regime  in which the tracks of CA06  are in agreement with
those employed in the present  paper embraces the location of \pr\ and
\pt, which have $\log T_{\rm eff}= 4.97$ and $\log T_{\rm eff}= 4.93$,
respectively.  Because  of  this,  it  is  expected  that  the  global
pulsation properties (i.e., asymptotic  period spacing, average of the
computed period spacing) of both  sets of models in that regime should
nearly agree,  and consequently the  asteroseismological inferences on
\pr\ and \pt\  based on these two different sets  of models should not
be substantially distinct.

In  Table \ref{compa}  we  present a  comparison  between the  stellar
masses inferred in CA06 (their Table 1) and in the present study. Note
that,  not surprisingly,  the stellar  masses derived  in CA06  are in
excellent  agreement with  the values  obtained in  the  present work,
irrespective of the particular  method employed, being the differences
in all of the cases below $2 \%$.  This is a clear indication that the
asteroseismological results  of this paper  and that of CA06  for \pr\
and \pt\  are not seriously  affected by the differences  between both
sets  of   models.  This  adds  credibility  and   robustness  to  the
asteroseismological  results  of the  present  study. This  conclusion
should change for the case of PG1159 pulsators with stellar masses too
departed  from the  value  $\sim 0.589  M_{\odot}$  and/or located  at
earlier evolutionary stages.

\begin{table}
\centering
\caption{Comparison between the stellar mass values (in  
$M_{\odot}$) obtained in CA06 and in the present work.}
\begin{tabular}{lccc}
\hline
\hline
\noalign{\smallskip}
\pr & $\Delta \Pi_{\ell}^{\rm a}$ & $\overline{\Delta \Pi_{\ell}}$ & 
Period fit \\
\noalign{\smallskip}
\hline
\noalign{\smallskip}
CA06 & $0.615$ & $0.575$ & $0.60$ \\
This work                   & $0.627$ & $0.578$ & $0.589$\\   
$\Delta M_*/M_*$                & 2 \%    & 0.3 \%  &  2 \%  \\
\hline
\hline
\noalign{\smallskip}
\pt & $\Delta \Pi_{\ell}^{\rm a}$ & $\overline{\Delta \Pi_{\ell}}$ & 
Period fit \\
\noalign{\smallskip}
\hline
\noalign{\smallskip}
CA06 & $0.595$ & $0.565$ & $0.55$ \\
This work                   & $0.597$ & $0.566$ & $0.542$\\   
$\Delta M_*/M_*$                & 0.3 \%  & 0.1 \%  & 1.5 \%  \\
\hline
\hline
\end{tabular}
\label{compa}
\end{table}

\section{Summary and conclusions}
\label{conclusions}

In  this  paper  we  presented  an asteroseismological  study  of  the
high-gravity, low-luminosity pulsating PG1159 stars \pr\ and \pt\, and
of  the high-luminosity  PNNV [WCE]  star  \ngc.  This  is the  fourth
article of a  short series of studies aimed  at exploring the internal
structure and evolutionary status of pulsating PG1159 stars which have
been intensively observed  through multi-site campaigns.  Our analysis
is  based on the  full PG1159  evolutionary models  of Althaus  et al.
(2005),  Miller  Bertolami \&  Althaus  (2006)  and  C\'orsico et  al.
(2006).   These  models  represent   a  solid  basis  to  analyze  the
evolutionary and pulsational properties of pre-white dwarf stars like
\pr, \pt, and \ngc.

We first  used the observed period-spacing data  to obtain estimations
of the stellar mass of \pr,  \pt, and \ngc. The results are summarized
in Table \ref{tabla-masas}, where we provide a summary of results from
the  present  and  the previous  works  by  us,  and also  from  other
pulsation and spectroscopic studies. We obtained three mass values for
each star: the first one by comparing the observed period spacing with
the asymptotic period spacing of our models (an inexpensive and widely
used  approach that  does not  involve pulsational  calculations); the
second one by comparing the  observed period spacings with the average
of the computed period spacing  (an approach that requires of detailed
period  computations);  and  the  third   one  on  the  basis  of  the
approximate formula  of Kawaler \&  Bradley (1994), which is  based on
the  behavior of  the asymptotic  period spacing  of a  large  grid of
quasi-evolutionary PG1159  models. The first and  the third approaches
are  almost equivalent,  and lead  to similar,  somewhat overestimated
values  of the stellar  mass for  \pr\ and  \pt. The  second approach,
clearly  more realistic,  conducts  to smaller  values  of $M_*$  (and
closer to the  spectroscopic inferences) in the case  of \pr\ and \pt,
and virtually the  same $M_*$ value than that  obtained from the first
approach in the case of \ngc.

In the  second part of  our work, we  sought for the models  that best
reproduce the  individual observed periods  of each star.   The period
fits were made on a grid of PG1159 models with a quite fine resolution
in effective  temperature ($\Delta T_{\rm eff}\sim  10-30$ K) although
admittedly coarse  in stellar  mass ($\Delta M_*  \sim 0.01 -  0.08 \,
M_{\odot}$).  We found asteroseismological models only for \pr\ and
\pt.   For \ngc\  we  were  unable  to  find a  clear and  unambiguous
seismological solution due to the existence of numerous and equivalent
minima characterizing the quality  function employed in the period-fit
procedure.  The pulsational properties of the ``best-fit'' models for
\pr\ and \pt\ are  summarized in Tables \ref{table1} and \ref{table2},
respectively.  In  particular, we predict  the values of the  rates of
period  change  to be  positive  and  in the  range  $(2  - 5)  \times
10^{-12}$ s/s. Unfortunately, we have been unable to check the reality
of this prediction because the  lack of any measurement of $\dot{\Pi}$
for \pr\ and  \pt\ for the moment.  The  structural characteristics of
these best-fit models are shown in Tables \ref{table3} and
\ref{table4}. In  particular, the  seismological masses are  closer to
the spectroscopic ones in light of our best-fit models, as we found in
our previous  works.  In these  tables we also show  the seismological
distances  and parallaxes  of \pr\  and  \pt.  We  found a  reasonable
agreement between our results and those of Kawaler et al. (1995), Reed
et al. (2000), Kawaler et al.  (2004), and CA06.  We stress that
almost all differences between our results (Sections \ref{char} and 
\ref{distance}) and those of earlier works are within the quoted errors.

In summary,  in this work  we have been  able to estimate  the stellar
mass  of  \pr, \pt,  and  \ngc\ on  the  basis  of the  period-spacing
information  alone.    We  have   also  been  successful   in  finding
asteroseismological  models for  \pr\ and  \pt\  from period-to-period
comparisons.  In  particular, the  $T_{\rm eff}$ and  $\log g$  of the
best-fit  models are  in very  good agreement  with  the spectroscopic
measurements.  Unfortunately, we  fail to found an asteroseismological
model for \ngc.  In principle,  this shortcoming of our study could be
indicating some inadequacy inherent  to the stellar modeling.  Another
possible alternative  could be  the fact that  the period  spectrum of
\ngc\  includes periodicities  associated with  $g$-modes  with radial
orders very  spaced from each other,  in such a way  that our $\chi^2$
procedure  of  period-fit  is  inefficient  to  isolate  a  clear  and
unambiguous asteroseismological solution. On  the other hand, it would
be kept in mind that while both \pr\ and \pt\ are classified as PG1159
stars, \ngc\ is  a [WCE] star. Although both  classes are suspected to
form an evolutionary sequence,  this possibility is still under debate
(Crowther  2008; Todt  et al.   2008) and  it could  not be  the case.
Therefore, the failure of our models to fit the period spectrum of
\ngc\ might be  indicative that this star has a very different 
evolutionary history than \pr\ and \pt.

We have also included a comparison between the models and results
of  the present  work  and those  of  the study  by  CA06. The  models
employed in the present work  are the result of the complete evolution
of the  progenitor stars, and as  a result, they  are characterized by
central  chemical abundances  and a  size of  the C-O  core  which are
consistent with  the value of the  stellar mass. This is  not the case
for the models employed by  CA06, which were artificially derived from
the  full  evolutionary  sequence  of $0.589  M_{\odot}$  computed  by
Althaus  et   al.  (2005).   In   spite  of  these   differences,  our
asteroseismological results  are in excellent agreement  with those of
CA06.   This adds  credibility and  robustness to  the results  of the
present study.

As the main conclusions of the present work, we can mention:

\begin{itemize}

\item[-] The full evolutionary models of PG1159 stars 
employed in  the present work lead to  asteroseismological results for
\pr\ and \pt\ that do not differ substantially from those predicted by
CA06 on the basis of evolutionary sequences generated artificially. We
note, however,  that this agreement between both  sets of computations
is valid for PG1159  stars located at the low-luminosity, high-gravity
regime after  the stars  have passed the  evolutionary knee in  the HR
diagram. It should be kept  in mind that this conclusion should change
for the case of stars located at earlier stages of evolution.

\item[-] At present, the PG1159 evolutionary models used in this 
work ---and in the previous  studies of this series--- remain the only
suitable for asteroseismological inferences  on stars that are located
at  the high-luminosity,  low-gravity regime  before  the evolutionary
knee, such as \rxj\ and \ngc.

\item[-] The detailed fitting of the individual periods 
(Sect.  \ref{fitting}) gives somewhat  different masses  than analysis
based on asymptotic period spacing (methods 1 and 3 of Sect.
\ref{period-spacing}), but in very good agreement with 
the values  of $M_*$ derived from  the average of  the computed period
spacings (method 2 of Sect. \ref{period-spacing}). Thus, method 2 is a
very appropriate way to estimating stellar masses, and detailed period
fits do  not significantly improve  the mass determinations.  We note,
however,  that the period-fit  approach yields  an asteroseismological
model from which one can  infer, in addition to $M_*$, the luminosity,
radius, gravity,  and distance  of the target  star. In  addition, the
period-fit  approach  does  not  require ---in  principle---  external
constraints such as the spectroscopic values of $T_{\rm eff}$ and $g$, 
i.e., the method works ``by letting the pulsation modes speak for 
themselves'' (see Metcalfe 2005 for an interesting discussion about 
this).

\item[-] The nonadiabatic stability analysis does not 
at the moment  predict instability for all of  the fitted modes.  This
means that, in the frame  of the linear nonadiabatic pulsation theory,
some pulsation modes detected in  \pr\ and \pt\ should not be excited.
It is not clear at this stage the origin of this discrepancy. Maybe it
could  be  attributed to  the  extreme  sensitivity  of the  stability
analysis of PG1159 stars to  the exact amounts of the main atmospheric
constituents (see Quirion et al. 2004 for details).

\item[-] The main conclusion of this series of papers ---the  
present work and C\'orsico et al.  (2007a, 2007b, 2008)--- is that for
most well-observed  pulsating PG1159 stars  (\rxj, \pg, \pp,  \pr, and
\pt) it is possible to  found a stellar model (the asteroseismological
model)   with  $M_*$   and  $T_{\rm   eff}$  near   the  spectroscopic
measurements to a  high internal accuracy. The next  step is of course
an assessment  of the question  if the asteroseismological  models can
provide more  accurate masses  for these objects.  The scatter  in the
masses  derived from  the different  asteroseismological  methods (see
Table \ref{tabla-masas})  suggests that  it may not  be the  case.  In
fact,  when  all   asteroseismological  methods  are  considered,  the
uncertainty in  the determination  of the mass  amounts to  $\sim 0.05
M_{\odot}$,  comparable  to  the  spectroscopic  one  ($\sim  0.05-0.1
M_{\odot}$; Werner  et al. 2008).   However, it is worth  noting that,
when results based  on asymptotic period spacing (an  approach that is
not correct for  the high-gravity regime of PG1159  stars; see Althaus
et al. 2008a) are not taken into account, the scattering in the derived
masses is of only $\sim 0.02 M_{\odot}$.

\end{itemize}

We  close the  paper by  noting  that the  PG1159 evolutionary  models
employed   in   our   series   of  asteroseismological   studies   are
characterized by thick He-rich  outer envelopes, as they are predicted
by the  standard theory  for the formation  of PG1159  stars. However,
Althaus et al.  (2008b) have recently demonstrated that the assumption
of  thinner  He-rich  envelopes  solves the  longstanding  discrepancy
between the measured  rates of period change in  the prototypical star
\pp\  and the  predictions of  theoretical  models.  In  view of  this
important result,  we are planning  future asteroseismological studies
for all the pulsating PG1159 stars analyzed in our series of articles,
but with non-canonical PG1159  models characterized by thinner He-rich
envelopes.


\begin{acknowledgements}
Part of this work was supported by the MEC
grant AYA05--08013--C03--01, by the  European Union FEDER funds, by
the  AGAUR, by AGENCIA through the Programa de Modernizaci\'on 
Tecnol\'ogica BID 1728/OC-AR, and by the PIP 6521 grant from CONICET. 
This  research has  made use  of NASA's  Astrophysics Data
System.
\end{acknowledgements}

\end{document}